# Mixed polarity reversible Peres gates

C. Moraga

Reversible Peres gates with more than two all over binary-valued control signals are discussed. Methods are disclosed for the low cost realization of this kind of Peres gates without requiring ancillary lines. It is shown that Peres gates with $n$ control signals may be obtained with a quantum cost of $2^{n+1} - n - 2$, using Feynman gates and controlled gates realizing the $\kappa$-th root of NOT, where $\kappa = 2^{n-1}$. Proper distribution of the controlled gates and their inverses allow driving the reversible Peres gate with control signals of different polarities.

*Introduction:* In the seminal work [1], the realization of minimal cost Toffoli gates [2] with several all over binary-valued control signals and no ancillary lines was presented. The design method was based on a Grey code to combine the control signals and activate or inhibit the elementary controlled gates realizing the $\kappa$-th root of NOT, where $\kappa = 2^{n-1}$ and NOT = [0 1; 1 0]. The method was shown to be scalable and the achievable quantum cost is $2^{n+1} - 3$. In [3] it was shown that based on the balanced nature of the Grey code it is possible to distribute the elementary controlled gates in such a way, that the Toffoli gate would become activated under every one of the $2^n$ possible binary control vectors. If the concept of polarity is borrowed from work on Reed Muller expressions it can be said that in [3] the method presented in [1] was extended to work with binary control signals of different polarities. The concept of mixed polarities for reversible circuits was possibly first introduced in [4].

The original Toffoli gate has two binary control signals, which are recovered at the output, and a target line where the product of the control signals is added modulo 2, to the input target signal. The extension to a gate with multiple all over binary-valued control signals is straight forward: all control signals should be recovered at the output, and on the target line the product of the control signals should be added modulo 2 to the input target signal. In the case of a Peres gate [5], the situation is different: only the first control signal is recovered, meanwhile the second output returns the modulo 2 addition of the control signals. At the target line, the behaviour of the Peres gate is as in the Toffoli gate. If additional control signals are considered for a Peres gate it is not obvious, which signals should be obtained at the new additional outputs. Only at the target line it seems clear that the behaviour should be as in the corresponding Toffoli gate, i.e., the product of the input control signals should be added modulo 2 to the input target signal. The following specification will be used in the present paper: A Peres gate with $n$ all over binary control signals $c_1, c_2,..., c_n$ and a target signal $t$ gives at the $i$-th output, the modulo 2 sum of the first $i$ control signals, $(1 \leq i \leq n)$, and $t \oplus c_1 c_2 ... c_n$ at the target output.

*Inductive Reasoning.*
In the case of $n = 2$, the symbol used for the Peres gate looks as the cascade of a Toffoli gate and a Feynman gate, as shown in Fig. 1a. A naïve interpretation at a level of a quantum realization based on Feynman and CV/CV† gates, as shown in Fig. 1b indicates that the external Feynman gate may cancel the internal one (both represented with dash lines and grey dots), finally leading to the realization of a Peres gate with a minimal quantum cost of 4 [1].

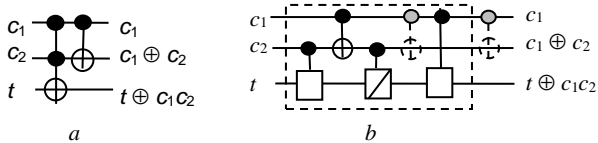

**Fig. 1:** *The classical Peres gate with two control signals.*
*a* Symbol and functionality of a Peres gate
*b* Naïve circuit interpretation of *a*, where the dotted block corresponds to a Toffoli gate. The white boxes represent CV-gates meanwhile the diagonalized box represents a CV† gate, which is the adjoint of CV.

The first intuition suggests that cascading appropriate Feynman gates to an optimal Toffoli gate with 3 or more binary control signals could lead to an optimal Peres gate. Using the method of [1], an optimal Toffoli gate with three control signals is obtained, as shown in Fig. 2, where the binary vectors denote the values of the coefficients $\alpha_1, \alpha_2, \alpha_3$ of the polynomial $\alpha_1 c_1 \oplus \alpha_2 c_2 \oplus \alpha_3 c_3$ which represents the driving functions of the controlled gates realizing the fourth root of NOT (white gate) or its adjoint (diagonalized gate). As mentioned in [1] the $\alpha$ coefficients of the driving functions follow a Grey code. (This may also be observed in the realization in Fig. 1b, with $\alpha$: [01] [11] [10]).

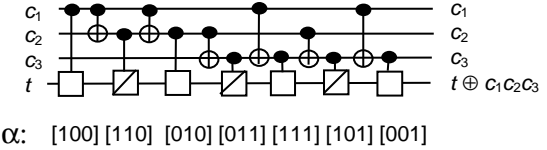

$\alpha$: [100] [110] [010] [011] [111] [101] [001]

**Fig. 2** *Realization of a Toffoli gate with three control signals using the method of Barenco et al.* [1].

The Toffoli gate of Fig. 2 has a quantum cost of 13 [1], which is optimal. In order to transform the control variables (recovered at the output) into the polynomials specified for the Peres gate (in Definition 1), two Feynman gates may be used, as shown in Fig. 3. It is fairly obvious that this circuit is scalable and has a quantum cost of $n$-1. However it is simple to see, that cascading the circuits of Figs. 2 and 3 would not lead to a cancellation of gates, because the intermediate signals are being used to drive gates on the target line. Quite on the contrary, the quantum cost of the resulting Peres gate would be 15, i.e., higher than that of the Toffoli gate. This contradicts what happens when $n = 2$.

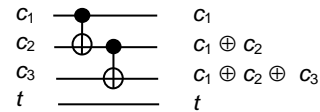

**Fig. 3** *Circuit to convert a Toffoli gate into a Peres gate, when* $n = 3$.

A simple observation of the (resulting) Peres gate shown in Fig. 1b makes clear that a reordering of the elementary controlled components may be done as illustrated in Fig. 4. An analysis of the $\alpha$ vectors allows the interpretation that they represent the coding of the (first three) natural numbers in bit-reversal order. Moreover, it may be noticed that if the Hamming weight of an $\alpha$ vector is odd, the controlled gates are CV and when it is even, the controlled gate is a CV†. This is consistent with the behaviour of the Peres gate: when both $c_1$ and $c_2$ are 1, the CV gates will be activated giving the expected negation and the CV† gate is inhibited behaving as an identity. The Peres gate then outputs NOT($t$) = $t \oplus 1 = t \oplus c_1 c_2$. If any one of the control signals is 0, The corresponding CV gate will be inhibited, meanwhile both the other CV gate and the CV† gate will be active, producing a global identity.

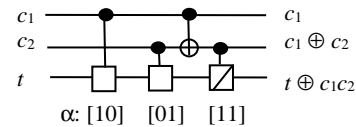

$\alpha$: [10] [01] [11]

**Fig. 4** *A different view of the Peres gate with reordered CV components*

Fig. 5 shows the growing of an intended Peres gate from $n = 2$ to $n = 4$ by using the new coding. The dash-framed sub-circuit has the same structure as in Fig. 4. The dash-dot-framed sub-circuit should realize a Peres gate with $n = 3$. Let $\kappa = 2^{n-1}$ then the controlled gates on the target line correspond to the $\kappa$-th root of NOT and their adjoints (see e.g. [6]). If the inputs to the circuit in Fig. 5 are $c_1, c_2, c_3, c_4$, and $t$, it is simple to see that the outputs will be $c_1, c_1 \oplus c_2, c_1 \oplus c_2 \oplus c_3, c_1 \oplus c_2 \oplus c_3 \oplus c_4$, and $t \oplus T$, where T will be given below. For space reasons, the $\alpha$ coefficients will be shown in a table format within Fig. 5.

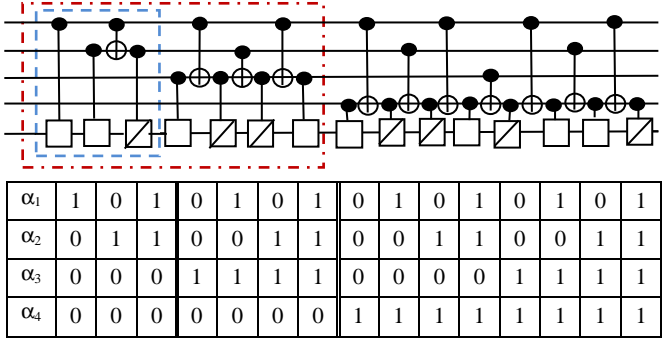

| $\alpha_1$ | 1 | 0 | 1 | 0 | 1 | 0 | 1 | 0 | 1 | 0 | 1 | 0 | 1 | 0 | 1 |
| --- | --- | --- | --- | --- | --- | --- | --- | --- | --- | --- | --- | --- | --- | --- | --- |
| $\alpha_2$ | 0 | 1 | 1 | 0 | 0 | 1 | 1 | 0 | 0 | 1 | 1 | 0 | 0 | 1 | 1 |
| $\alpha_3$ | 0 | 0 | 0 | 1 | 1 | 1 | 1 | 0 | 0 | 0 | 0 | 1 | 1 | 1 | 1 |
| $\alpha_4$ | 0 | 0 | 0 | 0 | 0 | 0 | 0 | 1 | 1 | 1 | 1 | 1 | 1 | 1 | 1 |

**Fig. 5** *Peres gate with 4 control signals based on the coding of the natural numbers with bit reversal order (shown as table for space reasons.)*

Proof of correctness. Let X denote the 8-th root of NOT. Recall that if an $\alpha$ vector has an even Hamming weight, the driven controlled gate will be CX†. Then if $c_1 = c_2 = c_3 = c_4 = 1$, all CX† gates will be driven by 0 and therefore will be inhibited, meanwhile all other controlled gates, CX gates, will be driven by 1, hence they will be active. Their cascade amounts to their 8-th power, therefore delivering NOT($t$) at the target line, i.e. T = $c_1 c_2 c_3 c_4$. W.l.o.g. assume that $c_1 = c_2 = c_3 = 1$, but $c_4 = 0$. This is equivalent to setting $\alpha_4 = 0$ all over its row. This will not affect the left part of the circuit, but at the right hand side, all Hamming weights will change their polarity and this means that all four CX† gates will be active, meanwhile the CX gates will be inhibited. At the same time, at the left part of the circuit four CX gates are active meanwhile the CX† gates are inhibited. The four active CX gates of the left part and the four CX† gates of the right part cancel each other and return an identity. It is not difficult to show that an identity is also generated with other combinations of 0-valued control signals. Therefore only if $c_1 c_2 c_3 c_4 = 1$ the circuit is active and outputs at the target $t \oplus 1$, i.e. it behaves as a controlled negation.

The structure grows incrementally (with $n$). For every new control variable, a new line is included that does not interfere with the gate with $n$-1 control variables. The new line acts as local target to collect the driving functions for the new controlled elementary gates. (In the table of $\alpha$ coefficients it may be seen that the corresponding half of the row has only 1-entries, and the patterns of $\alpha$-values in the upper rows are the same as in the former block(s).) When a new control variable is introduced, the only change to be done on the already existing part with $n$-1 control variables is the replacement of the elementary controlled gates by gates representing the next higher $\kappa$-th root of NOT.

A Peres gate with $n$ control variables has $2^n$ -1 elementary controlled gates. $n$ of the elementary controlled gates are driven directly by one control signal. The remaining $2^n$ -1 – $n$ elementary controlled gates require a Feynman gate (with quantum cost of 1) to generate each of the corresponding driving functions. Therefore the total quantum cost of the Peres gate amounts to $n + 2(2^n$ -1 – $n) = 2^{n+1} - n - 2$. Recall the circuit shown in Fig. 3 (for $n = 2$) to convert a Toffoli gate into a Peres gate. If the elementary gates are placed in reverse order –(if the circuit is seen "from right to left")– it will convert a Peres gate into a Toffoli gate, and it can be straight forward extended to higher values of $n$, with a quantum cost of $n – 1$. If this reversed circuit is added in cascade to the above discussed Peres gate, then a Toffoli gate (for the same $n$) would be obtained, and its quantum cost would be $(2^{n+1} - n - 2) + (n-1) = 2^{n+1} - 3$, which is the same minimal cost reported in [1].

*Mixed Polarity control*
Let [$c_1 c_2 ... c_n$] be a vector of control signals. Moreover let [$p_1 p_2 ... p_n$] be a polarity vector where $p_i \in \{0, 1\}$ for $1 < i < n$. Then [$c_1 \oplus p_1$  $c_2 \oplus p_2$ ... $c_n \oplus p_n$] is a control vector with the polarity specified by the polarity vector. The original control vector is sometimes called a 0-polarity control vector. For reversible gates, the standard control vector is [1 1 ... 1], and there are $2^n$-1 (non-zero) polarities. Recall that the functions driving the controlled elementary gates realizing the $\kappa$-th root of NOT or their adjoints, have the structure $\alpha_1 c_1 \oplus \alpha_2 c_2 \oplus .... \oplus \alpha_n c_n$, which, if a polarity of the control vector is considered, turns into $\alpha_1 (c_1 \oplus p_1) \oplus \alpha_2 (c_2 \oplus p_2) \oplus .... \oplus \alpha_n (c_n \oplus p_n)$. If the standard control vector is taken as reference, then it is simple to see that the resulting control vector has as components, the complement of the corresponding components of the polarity vector.

Let it be assumed that a Peres gate with $n$ control signals has been designed using the binary coding of the natural numbers to determine the $\alpha$ coefficients of the driving functions of the elementary gates. If a polarity is considered, changing $c_n$ into its complement, (i.e. $c_n = 0$) then all entries of the last row may be replaced by 0.This has the effect of "subtracting 1" mod 2 (which is equivalent to adding 1 mod 2) to the value of the driving functions in all the columns of the right hand side, i.e., taking their complements, and, consequently, the effect of replacing the elementary controlled gates by their corresponding adjoints, to activate the Peres gate only if the control vector is 11...10. Since all other polarities may be obtained by changing one control signal at a time, Peres gates of minimal quantum cost, controlled by any non-zero control vector may be obtained by applying iteratively the former procedure, as discussed in [3] for Toffoli gates. (A "parallel" version would start with a Peres gate designed for the standard control vector. For any other control vector, the value of the driving functions should be calculated. Whenever this value is 1, the elementary controlled gate should realize the $\kappa$-th root of NOT; otherwise, its adjoint, which is its inverse, should be taken.)

If all control signals were set to 0, the former Peres gate would be inhibited. If however *all* elementary controlled gates realize the $\kappa$-th root of NOT, for any other control vector, it is simple to show that the gate will behave as NOT, i.e. the target output will be $t \oplus$ ($c_1 \vee c_2 \vee ... \vee c_n$): the Peres gate behaves as an OR gate. However, according to the De Morgan laws, the complement of the OR of ($c_1, c_2, ..., c_n$) equals the AND of the complemented arguments, therefore, if the target signal $t$ may be complemented (either by initialization or by adding an inverter), a Peres gate with a 0-control vector is obtained.

*Conclusion:* A method has been disclosed that allows the design of *both* Peres *and* Toffoli gates with any number of over all binary control signals and mixed polarities, with a quantum cost of $2^{n+1} - n - 2$ and $2^{n+1} - 3$, respectively, without ancillary lines. The obtained Toffoli gates have the same minimal cost as reported in [1] and [3], but have a different distribution of the elementary controlled gates. The scalability of the method may clearly be seen in the structure of the resulting circuits.

*Acknowledgments:* This work was partially supported by the CICYT Spain, under project TIN 2011-29827-C02-01.

C. Moraga (*European Centre for Soft Computing*. 33600 Mieres, Spain)
E-mail: claudio.moraga@softcomputing.es
C. Moraga: also with *Faculty of Computer Science, TU Dortmund University*, Germany.